\documentclass[cits]{PoS}

\title{Impact of SUSY-QCD corrections on neutralino-stop co-annihilation and the neutralino relic density}

\ShortTitle{Impact of SUSY-QCD corrections on neutralino-stop co-annihilation and the relic density}

\author{\speaker{Julia Harz}\\
        Deutsches Elektronen-Synchrotron (DESY)\\  Notkestra{\ss}e 85, D-22607 Hamburg, Germany\\
        E-mail: \email{julia.harz@desy.de}}

\author{Bj\"orn Herrmann\\
       LAPTh, Universit\'e de Savoie\,/\,CNRS, 9 Chemin de Bellevue\\
       B.P.\ 110, F-74941 Annecy-le-Vieux, France
       }

\author{Michael Klasen\\
       Institut f\"ur Theoretische Physik, Westf\"alische Wilhelms-Universit\"at M\"unster\\
       Wilhelm-Klemm-Stra{\ss}e 9, D-48149 M\"unster, Germany
       }

\author{Karol Kova\v{r}\'{\i}k\\
       Institute for Theoretical Physics, Karlsruhe Institute of Technology\\
       D-76128 Karlsruhe, Germany
       }

\author{Quentin Le Boulc'h\\
       Laboratoire de Physique Subatomique et de Cosmologie,\\
       Universit\'e Joseph Fourier/CNRS-IN2P3/INPG\\
       53 Rue des Martyrs, F-38026 Grenoble, France
       }

\abstract{We have calculated the full O($\alpha_s$) supersymmetric QCD corrections to neutralino-stop co-annihilation into electroweak vector and Higgs bosons within the Minimal Supersymmetric Standard Model (MSSM). We performed a parameter study within the phenomenological MSSM and demonstrated that the studied co-annihilation processes are phenomenologically relevant, especially in the context of a 126 GeV Higgs-like particle. By means of an example scenario we discuss the effect of the full next-to-leading order corrections on the co-annihilation cross section and show their impact on the predicted neutralino relic density. We demonstrate that the impact of these corrections on the cosmologically preferred region of parameter space is larger than the current experimental uncertainty of WMAP data.}

\FullConference{ Proceedings of the Corfu Summer Institute 2012 \\
		 September 8-27, 2012\\
		 Corfu, Greece}

\newcommand{\MO}{{\tt micrOMEGAs}}
\newcommand{\DS}{{\tt DarkSUSY}}

\newcommand{\SPheno}{{\tt SPheno}}

\newcommand{\FeynArts}{{\tt FeynArts}}
\newcommand{\FORM}{{\tt FORM}}
\newcommand{\Fortran}{{\tt Fortran}}

\newcommand{\FeynCalc}{{\tt FeynCalc}}

\newcommand{\GeV}{{\;\mathrm{GeV}}}

\newcommand{\dd}{{\rm d}}

\begin{document}

\section{Introduction}%
Different cosmological observations provide evidence for the existence of Cold Dark Matter (CDM). Through measurements of the WMAP satellite in combination with baryonic acoustic oscillation and supernova data \cite{WMAP7} the relic density of dark matter can be constrained very precisely to the interval
\begin{equation}
	\Omega_{\rm CDM}h^2 ~=~ 0.1126 ~\pm~ 0.0036
	\label{Eq:WMAP}
\end{equation}%
at $1\sigma$ confidence level, where $h$ in units of 100 km s$^{-1}$\,Mpc$^{-1}$ denotes the present Hubble expansion rate. One possible candidate for dark matter is the so called weakly interacting massive particle (WIMP). However, the Standard Model of particle physics does not provide such a particle fulfilling all relevant constraints. In contrast, the Minimal Supersymmetric Standard Model (MSSM) with conserved $R$-parity, contains an appropriate dark matter candidate, the lightest neutralino $\tilde{\chi}^0_1$, a stable WIMP.\\
To calculate the dark matter relic density, the time evolution of the dark matter number density $n_{\chi}$, described by the Boltzmann equation, has to be solved. As we want to consider the case when heavier, unstable supersymmetric particles survive in the Universe for sufficient time to affect the dark matter relic density, we have to account for all interactions between two surviving particles $i$ and $j$. This can be described by an effective form \cite{GriestSeckel, EdsjoGondolo}
\begin{equation}\label{Eq:Boltzmann}
 \frac{\dd n_{\chi}}{\dd t} ~=~ - 3 H n_{\chi} - \langle \sigma_{\rm eff}v \rangle
 \left[ n_{\chi}^2 - (n_{\chi}^{\rm eq})^2 \right].
\end{equation}%
As all particles finally decay into the dark matter particle, the total number density can be written as $n_\chi=\sum_i n_i$ with $n_{i}$ being the number density for each particle species.
The effective (co-)annihilation cross section is given by
\begin{equation}
 \langle \sigma_{\rm eff}v \rangle ~=~ \sum_{i,j} \sigma_{ij}v_{ij}
 \frac{n^{\rm eq}_i}{n_{\chi}^{\rm eq}} \frac{n^{\rm eq}_j}{n_{\chi}^{\rm eq}}\,,
 \label{Eq:sigmaeff}
\end{equation}%
where the sum runs over all MSSM particles $i$ and $j$. The relative velocity of the two interacting particles is described by $v_{ij}$. The ratio between their respective number density in thermal equilibrium $n^{\rm eq}_i$ and the number density of the dark matter particle $n_{\chi}^{\rm eq}$ at temperature $T$, is Boltzmann suppressed
\begin{equation}\label{Eq:BoltzmannSupression}
 \frac{n^{\rm eq}_i}{n_{\chi}^{\rm eq}} ~\sim~ \exp\left[ - \frac{m_i-m_{\chi}}{T}
 \right].
\end{equation}%
Only if another particle is almost mass degenerate with the dark matter particle, the neutralino in our case, the effective cross section gets a significant contribution of the corresponding processes. Thus annihilation of two neutralinos, co-annihilation of a neutralino with other gauginos, or co-annihilation of neutralinos with light sleptons or squarks can contribute in a sizeable manner.\\
Having solved the Boltzmann equation numerically, the neutralino relic density can be calculated by
\begin{equation}
 \label{Eq:OmegaMass}
 \Omega_{\chi}h^2 ~=~ \frac{m_{\chi} n_{\chi}}{\rho_{\rm crit}}\,,
\end{equation}%
where $m_{\chi}$ is the mass and $n_{\chi}$ the current number density of the neutralino and $\rho_{\rm crit}$ the critical density of the Universe.\\
Comparing the theoretically predicted dark matter relic density with the experimentally obtained limits allows one to constrain the MSSM parameter space and thus to combine information of cosmological observations with collider searches and precision measurements.\\
However, the theoretical prediction of the neutralino relic density suffers from different sources of uncertainties, which arise from cosmology as well as from particle physics, for example. From the cosmological point of view the uncertainties are connected to the choice of the cosmological model \cite{Hamann} or the definition of the Hubble expansion rate before Big Bang Nucleosynthesis \cite{Arbey}. On the particle physics side, uncertainties arise due to the calculation of essential parameters like physical masses or couplings of supersymmetric particles. A possible source for these discrepancies is for example the different treatment of radiative corrections or a different implementation of renormalization group equations in spectrum calculators \cite{Belanger}.\\
Another uncertainty, the one we address in the following, concerns the actual precision of the calculation of the (co-)annihilation cross sections. Current public dark matter tools like {\DS} \cite{DarkSusy} or {\MO} \cite{micrOMEGAs2007} evaluate the relic density on the basis of (co-)annihilation cross sections calculated at an effective tree-level. Especially with the even more precise data which will be provided by PLANCK in the very near future, the accuracy of the theoretical prediction has to be improved to meet the experimental precision. The non-negligible impact of next-to-leading order corrections on the dark matter relic density has been discussed in several previous analyses. SUSY-QCD corrections to neutralino pair annihilation into a quark-antiquark pair have been studied in Refs. \cite{DMNLO_AFunnel, DMNLO_mSUGRA, DMNLO_NUHM}. The corresponding electroweak corrections have been addressed in Refs. \cite{Sloops2007, Sloops2009, Sloops2010}, the co-annihilation of a neutralino with another gaugino has been also discussed in Refs. \cite{Sloops2009, Sloops2010}. Further studies regarding neutralino pair annihilation or co-annihilation with a tau slepton, have been performed on the basis of an effective coupling approach in Refs. \cite{Sloops2011,EffCouplings}. SUSY-QCD corrections to co-annihilation of a neutralino with a stop have been only considered in Ref. \cite{Freitas2007} so far. Although studying the very specific cases of co-annihilation of a bino-like neutralino with a right-handed stop into a top quark and a gluon or into a bottom quark and a $W$-boson, a significant impact on the dark matter relic density was shown. Similar results were found in the aforementioned studies.\\
However, depending on the region of parameter space, other final states, including those with other electroweak gauge and Higgs bosons, can become dominant. Thus we extend the analysis of QCD and SUSY-QCD corrections to co-annihilation of a neutralino with a stop by computing the general case of neutralino-stop co-annihilation into a quark and a Higgs or an electroweak vector boson.\\
In the following, the phenomenology of neutralino-stop co-annihilation in the MSSM is discussed. Thereafter the calculation of the next-to-leading order corrections to the relevant processes is described and their impact on the co-annihilation cross sections and the dark matter relic density is shown.

\section{Phenomenology of neutralino-stop co-annihilation}
\subsection{Neutralino-stop co-annihilation in the context of a 126 GeV Higgs boson}
\begin{figure*}[t] \centering
 \includegraphics[scale=0.57]{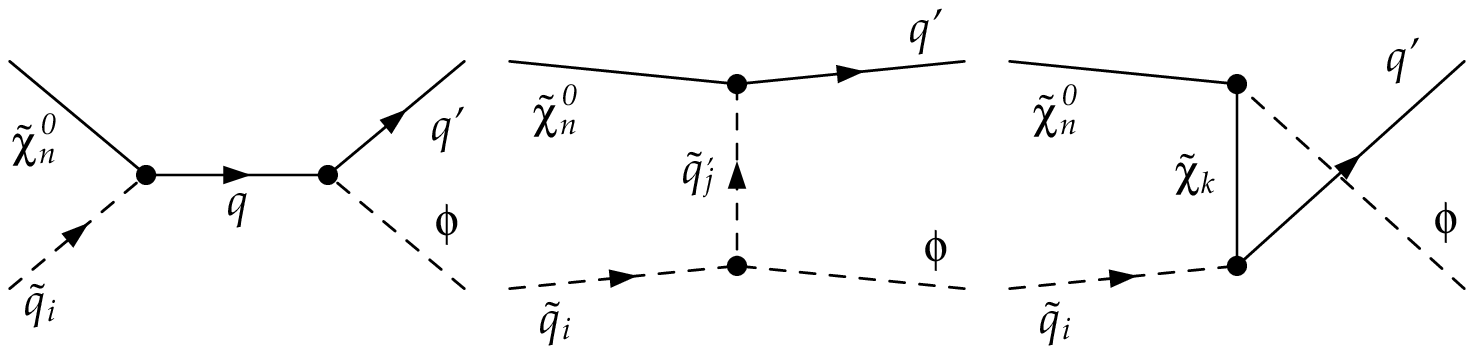}
 \includegraphics[scale=0.57]{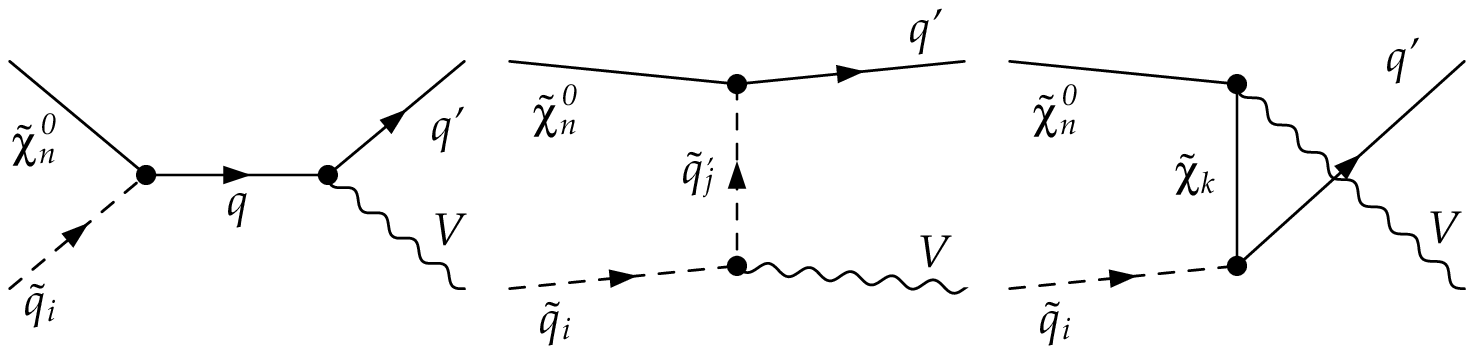}
 \caption{Leading-order Feynman diagrams for neutralino-squark
 co-annihilation into a quark and a Higgs boson ($\phi=h^0,H^0,A^0,H^{\pm}$)
 or an electroweak gauge boson ($V=\gamma,Z^0,W^{\pm}$). The $u$-channel is absent for a photon in the final state.}
 \label{Fig:TreeDiagrams}
\end{figure*}

In certain regions of the MSSM parameter space co-annihilation of the lightest neutralino with the next-to-lightest supersymmetric particle (NLSP) can become dominant in comparison to other (co-)annihilation processes. A particularly interesting example of such a NLSP is the stop: Especially when the trilinear coupling $A_t$ gets large absolute values, its chirality eigenstates can mix significantly. The lighter mass eigenstate can then be almost mass-degenerate with the lightest neutralino, which induces the studied neutralino-stop co-annihilation \cite{StopCoann1}.\\
Interpreting the recent observation of a boson with a mass of about 126 GeV \cite{ATLAS2012, CMS2012} as a light CP-even Higgs boson ($h^0$), a specific choice of parameters in the stop and sbottom sector within the MSSM is implied \cite{Arbey2012}. The reason lies in the interplay of these parameters regarding the lightest Higgs boson mass. The leading contribution to its mass arises from a loop containing stops, which can together with the tree-level be expressed as follows \cite{Haber1996, Badziak2012}
\begin{equation}
 m_{h^0}^2 ~=~ m_Z^2 \cos^2 2\beta + \frac{3 g^2 m_t^4}{8 \pi^2 m_W^2} 
 \left[ \log\frac{M_{\rm SUSY}^2}{m_t^2} + \frac{X^2_t}{M^2_{\rm SUSY}}
 \left( 1 - \frac{X^2_t}{12 \,M^2_{\rm SUSY}} \right) \right]\,,
 \label{Eq:HiggsMass}
\end{equation}%
where $X_t = A_t - \mu / \tan\beta$ and $M_{\rm SUSY} = \sqrt{m_{\tilde{t}_1} m_{\tilde{t}_2}}$. The maximal contribution from stop mixing is obtained for $|X_t| \sim \sqrt{6} M_{\rm SUSY}$, which favors a sizable trilinear coupling $A_t$ and consequently a rather light stop.\\
Besides, a light third generation of sfermions is helpful in order to reduce fine-tuning and to stay compatible with experimental constraints at the same time, like for example in ``natural'' SUSY models \cite{naturalSUSY1, naturalSUSY2}. 
In the following, we focus on neutralino-stop co-annihilation with a quark and an electroweak vector or Higgs boson in the final state. The corresponding leading order Feynman diagrams are shown in Fig.~\ref{Fig:TreeDiagrams}. At tree level, the process is mediated either by an $s$-channel quark, a $t$-channel squark, or a $u$-channel neutralino or chargino exchange.\\

\subsection{Parameter study within the pMSSM}
\begin{figure*}[t]
\begin{center}
\includegraphics[scale=0.76]{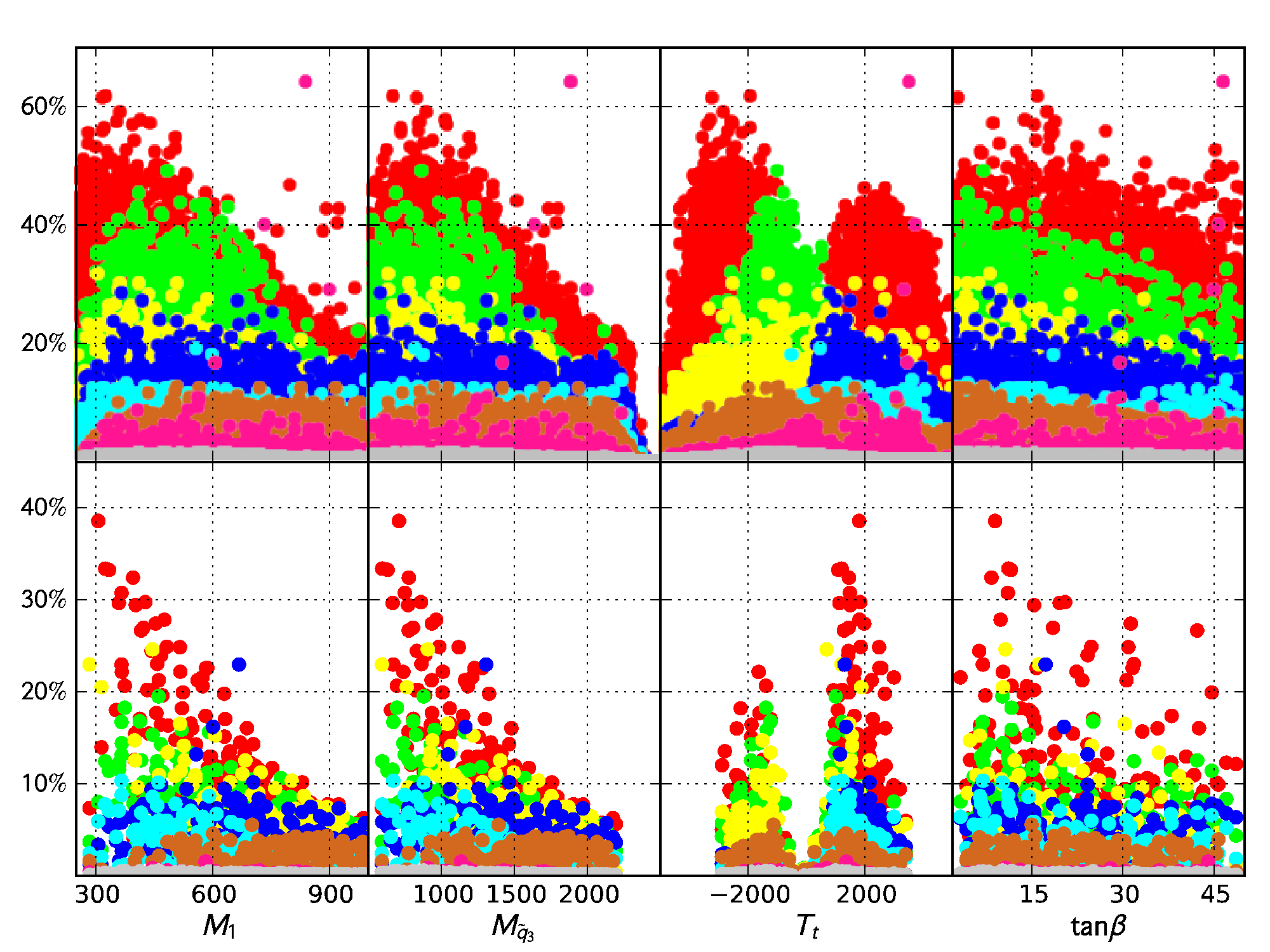}
\end{center}
\caption{Relative contributions of the neutralino-stop co-annihilation
 channels for the generated parameter points as a function of the input
 parameters $M_1$, $M_{\tilde{q}_3}$, $T_t$, and $\tan\beta$ before (top) and
 after (bottom) applying the selection cuts of Eq.\ (\protect\ref{Eq:SelectionCuts}).
 Shown are the contributions from $th^0$ (red), $tg$ (green), $tZ^0$ (blue),
 $tH^0$ (yellow), $bW^+$ (cyan), $tA^0$ (brown), $bH^+$ (pink), and
 $t\gamma$ (gray) final states. The parameters $M_1$, $M_{\tilde{q}_3}$, and
 $T_t$ are given in GeV.}
\label{Fig:RandomScan}
\end{figure*}

In order to quantify the relative importance of the processes in Fig.~\ref{Fig:TreeDiagrams}, we have performed a random scan in the phenomenological MSSM. For simplification we restricted the model to the following eight parameters: The slepton sector is characterized by one single mass parameter $M_{\tilde{\ell}}$, the squark sector is described by $M_{\tilde{q}_{1,2}}$ for the first and second generation squarks, and the common mass parameter $M_{\tilde{q}_3}$ for the third generation squarks. Except for $A_t$ in the stop sector, all trilinear couplings are set to zero. In our study we parametrize the trilinear coupling as $T_t = Y_t A_t$ with $Y_t$ being the top Yukawa coupling. With the wino and gluino masses fixed by $2 M_1 = M_2 = M_3/3$, which is motivated by gaugino mass unification at the GUT scale, all gaugino masses are defined through the bino mass parameter $M_1$. The Higgs sector is defined by the higgsino mass parameter $\mu$, the pole mass of the pseudoscalar Higgs boson $m_A$ and the ratio $\tan\beta$ of the two vacuum expectation values of the Higgs doublets. According to the SPA-convention \cite{SPA2005} these soft-breaking parameters are defined at the scale $Q=1$ TeV. We have randomly generated 1.2 million parameter sets with the eight input parameters lying in the following ranges:
\begin{minipage}[h][3.2cm][t]{0.5\textwidth}
\begin{eqnarray}
 500~{\rm GeV} \leq M_{\tilde{q}_{1,2}} &\leq&~ 4000~{\rm GeV}, \nonumber\\
 100~{\rm GeV} \leq M_{\tilde{q}_3}     &\leq&~ 2500~{\rm GeV} , \nonumber\\
 500~{\rm GeV} \leq M_{\tilde{\ell}}   ~&\leq&~ 4000~{\rm GeV}, \nonumber \\
                    |T_t|              ~&\leq&~ 5000~{\rm GeV} ,  \nonumber
\end{eqnarray}
\end{minipage}
\begin{minipage}[h][3.2cm][t]{0.5\textwidth}
\begin{eqnarray}
 200~{\rm GeV} \leq M_1                ~&\leq&~ 1000~{\rm GeV}, \nonumber\\
 100~{\rm GeV} \leq m_A                ~&\leq&~ 2000~{\rm GeV} ,  \\
                            |\mu|      ~&\leq&~ 3000~{\rm GeV} , \nonumber \\
 2             \leq        \tan\beta    &\leq&  50 . \nonumber
\end{eqnarray}
\end{minipage}\\
Using {\SPheno} \cite{SPheno} (version {\tt 3.2.1}) the mass spectrum and mixing matrices have been determined for each point in the parameter space. The contributions of the individual (co-)annihilation channels as well as the neutralino relic density $\Omega_{\chi}h^2$ have been evaluated using {\MO} \cite{micrOMEGAs2007} (version {\tt 2.4.1}). The necessary numerical values of the Standard Model parameters have been taken from Ref.\ \cite{PDG2012}.\\
In Fig.~\ref{Fig:RandomScan} the relative contribution of the different neutralino-stop co-annihilation channels for the generated parameter points as a function of the most relevant input parameters is shown. As can be seen, in a large number of parameters sets co-annihilation of a neutralino with a stop contributes significantly to the overall (co-)annihilation cross section.\\
In order to focus only on experimentally viable scenarios, we impose the following selection cuts on the data:
\begin{eqnarray}
 0.0946 &\leq~  \Omega_{\chi}h^2 ~\leq& 0.1306 , \nonumber \\ 
 120\,{\GeV} &\leq~ m_{h^0} ~\leq& 130\,{\GeV} , \\  \label{Eq:SelectionCuts}
 2.77 \cdot 10^{-4} &\leq~ {\rm BR}(b\to s\gamma) ~\leq& 4.33\cdot 10^{-4} .\nonumber
\end{eqnarray}%
The first selection cut chooses the parameter sets where the relic density lies within a $5~\sigma$ confidence interval. Considering a theoretical uncertainty of about 3 GeV, the second one corresponds to a medium cut on the lightest Higgs boson mass. The third selection cut is chosen to be a $3\sigma$ interval around the observed value of ${\rm BR}(b\to s\gamma)=(3.55 \pm 0.26)\cdot 10^{-4}$ \cite{HFAG}.
The relative contributions of the studied co-annihilation channels of the remaining parameter sets after imposing the cuts are shown in the lower part of Fig.~\ref{Fig:RandomScan}.\\
Comparing the data points before and after applying the cuts, several interesting features can be observed. The shape of the distribution is not significantly changed, but the density of the points is reduced. The main characteristics of the co-annihilation, the degenerate masses of the lightest neutralino and the stop, is reflected in the left and left-center column of Fig.~\ref{Fig:RandomScan}, where the dependence of the relative contribution on the the gaugino mass parameter $M_1$ and the third-generation squark mass parameter $M_{\tilde{q}_3}$ is depicted. For large values of both input parameters, co-annihilations cease to be important and annihilations of stops take their place as the dominating contribution of the total cross section.\\
After imposing the selection cuts, particularly the channel with the top and the gluon in the final state is reduced, whereas the final state with the lightest Higgs remains the dominant contributing channel. The reason for this behavior becomes visible in the right-center column of Fig.~\ref{Fig:RandomScan} where the dependence on the trilinear coupling $T_t$ is depicted. To match the experimental observations, a sizable trilinear coupling $T_t$ is preferred, which agrees with our previous discussion of Eq.\ (\ref{Eq:HiggsMass}). Moreover, positive values for $T_t$ are slightly favored, since they allow a better maximization of the Higgs boson mass \cite{Arbey2012}. This results also in an enhancement of the Higgs-squark-squark coupling, which is present in the $t$-channel of the $t h^0$ final state and leads to the dominating relative contribution of the co-annihilation into Higgs final states. This means that the same mechanism which drives the lightest Higgs boson mass through important stop-loop contributions to the observed value, is also responsible for the increase of neutralino-stop co-annihilation into the lightest Higgs boson together with a top quark.\\
A similar behavior is found for the Higgs parameter $\mu$, which is not displayed: Larger values of $\mu$ enhances the Higgs-sfermion-sfermion coupling (mainly for the heavy CP-even Higgs), which has the same consequences as for large values of $T_t$.\\
In contrast, the dependence on $\tan\beta$ is less pronounced. The remaining input parameters, such as those related to first and second generation squarks, sbottoms, and sleptons, as well as the higgsinos are less important in this context and are therefore not displayed in Fig.~\ref{Fig:RandomScan}.\\
In the following we present an example for a characteristic scenario with dominant contribution of the $t h^0$ final state. The corresponding input parameters and masses are given in Tab.~\ref{Tab:Scenarios}. The value of the neutralino relic density and the relative contributions of the dominating channels are listed in Tab.~\ref{Tab:Channels}. For a more detailed discussion of further scenarios we refer the reader to our recent paper \cite{StopCoannOur1}.\\
\begin{table*}\centering
\scriptsize
 \begin{tabular}{|cccccccc|cccc|}
 \hline
  ~~~ $M_1$ ~~~& ~~~$M_{\tilde{q}_{1,2}}$~~~ & ~~~$M_{\tilde{q}_3}$~~~ & ~~~$M_{\tilde{\ell}}$~~~ & ~~~$T_t$~~~ & ~~~$m_{A}$~~~ &
 ~~~$\mu$~~~ & ~~~$\tan\beta$~~~ & ~~~$m_{\tilde{\chi}^0_1}$~~~ & ~~~$m_{\tilde{t}_1}$~~~ & ~~~$m_{h^0}$~~~ & ~~~$m_{H^0}$~~~  \\
 \hline
$306.9$ & $2037.7$ & $709.7$ & $1499.3$ & $1806.5$ & $1495.6$ & $2616.1$ & $9.0$ & 307.1 & 350.0 & 124.43 & 1530.72 \\
 \hline
 \end{tabular}
 \caption{Example for a characteristic scenario within the pMSSM. Given are the input parameters as described in the text and selected particle masses. All values
 except $\tan\beta$ are given in GeV.}
 \label{Tab:Scenarios}
\end{table*}%
\begin{table*}\centering
\scriptsize
 \begin{tabular}{|c|c|c|c|c|c|}
 \hline
   ~~$\Omega_{\chi}h^2$~~ & ~~ $\tilde{\chi}^0_1 \tilde{t}_1 \rightarrow t h^0$ ~~ & ~~ $\tilde{\chi}^0_1 \tilde{t}_1 \rightarrow b W^+$ ~~ & ~~ $\tilde{\chi}^0_1 \tilde{t}_1 \rightarrow t Z^0$ ~~ & ~~ Sum ~~ \\
 \hline
 0.114 & 38.5\% & 5.9\%  & 3.4\% & 47.8\% \\
 \hline	
 \end{tabular}
 \caption{Neutralino relic density and relative contributions of neutralino-stop
 co-annihilation into a quark and a Higgs or electroweak gauge boson for the
 example point of Tab.\ \protect\ref{Tab:Scenarios}.}
 \label{Tab:Channels}
\end{table*}%
The chosen scenario is characterized by the dominant neutralino-stop co-annihilation into a top quark and a light Higgs boson. This is due to the large value of $T_t$ in this parameter point, which enhances the Higgs-sfermion-sfermion coupling and thus the squark exchange in the $t$-channel. Final states including a top quark and a $Z$-boson as well as a bottom quark and a $W$-boson contribute to a clearly lesser extent. Summing up the contributions of all final states, neutralino-stop co-annihilation accounts to almost half of the total (co-)annihilation cross section at this point. This makes the chosen example point an interesting scenario for the following further studies.

\section{One-loop cross section}\label{Sec:Calculation}
\subsection{Corrections and divergence treatment}
\begin{figure*}[t]\centering
	\includegraphics[scale=1.0]{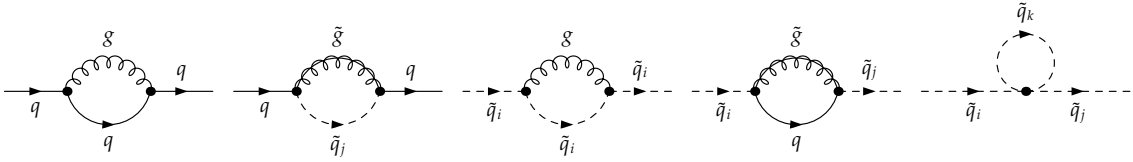}
	\caption{Self-energy corrections for quarks and squarks at one-loop level in QCD contributing to neutralino-squark co-annihilation.}
	\label{Fig:SelfEnergies}
\end{figure*}

\begin{figure*}[t]\centering
	\includegraphics[scale=0.77]{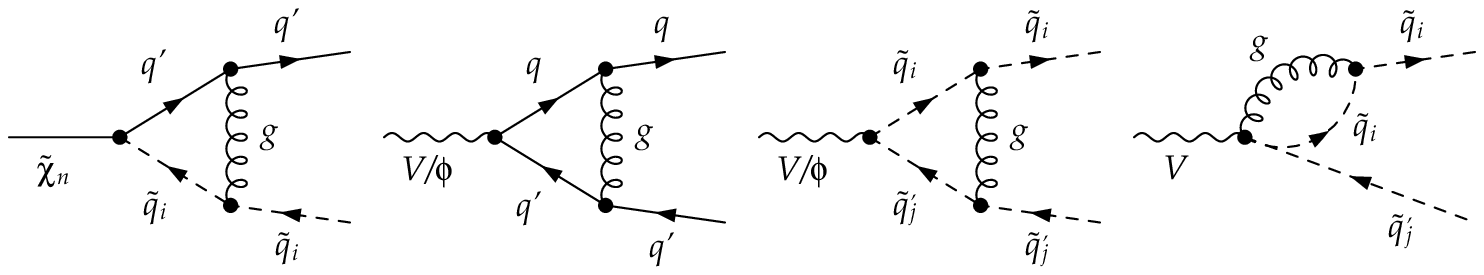}
	\includegraphics[scale=0.77]{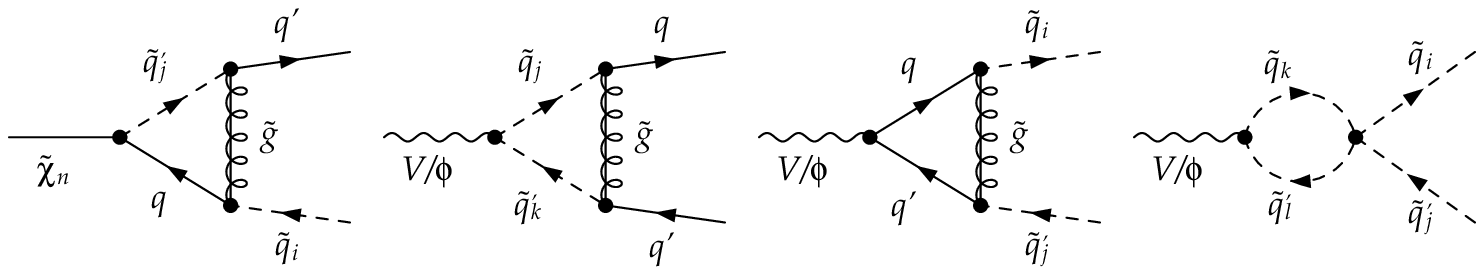}
	\caption{Vertex corrections at one-loop level contributing to neutralino-squark co-annihilation 
	into quark and Higgs ($\phi$) or electroweak vector ($V$) boson. The diagram involving the $V-g-\tilde{q}-\tilde{q}$ vertex is only present for the case of a vector boson in the final state.}
	\label{Fig:VertexCorrections}
\end{figure*}

\begin{figure*}[t]
	\includegraphics[scale=1.0]{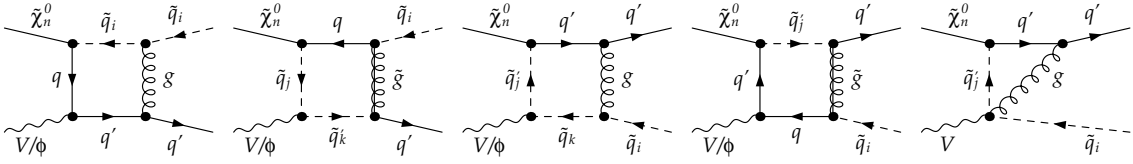}
	\caption{Four-point diagrams at one-loop level contributing to neutralino-squark co-annihilation into quark and Higgs ($\phi$) or electroweak vector ($V$) boson. The last diagram is absent for a scalar in the final state.}
	\label{Fig:BoxDiagrams}
\end{figure*}

\begin{figure*}[t]\centering
	\includegraphics[scale=1.0]{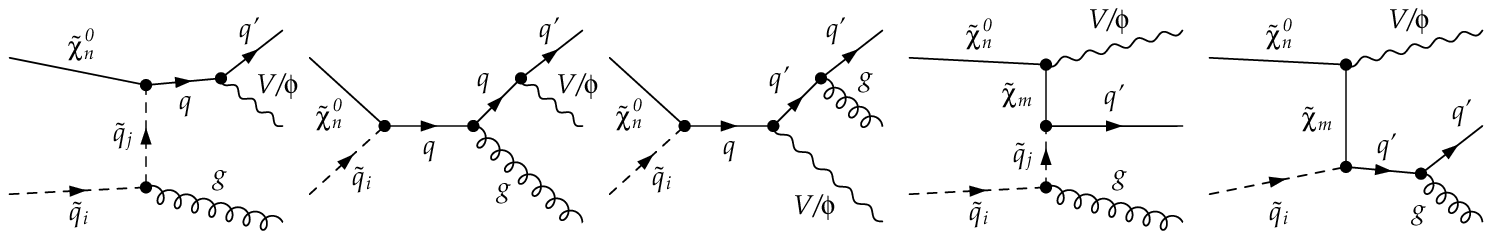}
	\includegraphics[scale=0.77]{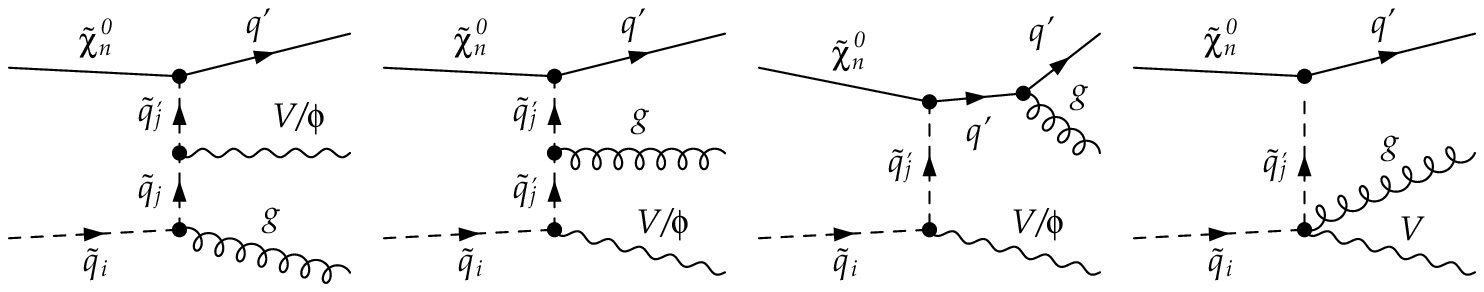}
	\caption{Real gluon emission diagrams at one-loop level contributing to neutralino-squark 
	co-annihilation into quark and Higgs ($\phi$) or electroweak vector ($V$) boson. The last diagram involving the four-vertex is absent for a scalar in the final state.}
	\label{Fig:RealEmission}
\end{figure*}

The virtual corrections for the co-annihilation processes, shown in Fig.~\ref{Fig:TreeDiagrams}, contain propagator corrections (see Fig.~\ref{Fig:SelfEnergies}), vertex corrections (see Fig.~\ref{Fig:VertexCorrections}) and box contributions (see Fig.~\ref{Fig:BoxDiagrams}). They have been calculated analytically and cross-checked using the publicly available tools {\FeynArts} \cite{FeynArts}, {\FeynCalc} \cite{FeynCalc}, and {\FORM} \cite{FORM}. The arising divergences are regularized by calculating in $D=4-2\varepsilon$ dimensions. To preserve supersymmetry the dimensional reduction regularization scheme ($\overline{\tt DR}$) has been applied. For canceling the ultraviolet (UV) singularities, corresponding counterterms to the relevant MSSM parameters and fields have been introduced. With the choice of a hybrid on-shell/$\overline{\tt DR}$ renormalization scheme, we minimize the sources for potential problems connected to sensitive parameters, e.g. the bottom trilinear coupling $A_b$. Using this scheme we can treat consistently the relevant parameters in the quark and squark sector of the MSSM for all (co-)annihilation processes over a large region of parameter space. The resulting expressions for the virtual corrections and their counterterms have been implemented in a numerical {\Fortran} code \cite{DMNLO}. We have explicitly verified that after renormalization all UV divergences cancel.\\
A special treatment is necessary for canceling the infrared (IR) divergences, which arise in the diagrams where a gluon is exchanged (see Figs.~\ref{Fig:VertexCorrections}, \ref{Fig:BoxDiagrams}). These divergences cancel against similar ones that come from the real radiation diagrams (see Fig.~\ref{Fig:RealEmission}) where a gluon is emitted off a quark or squark. In contrast to the UV divergences, the IR divergence treatment is slightly more difficult, as the divergences of the real emission diagrams come from the integration over the gluon phase-space. \\
Different approaches exist to deal properly with this kind of divergences. For our study we use the phase-space slicing method \cite{GieleGlover,HarrisOwens,Denner:1991kt} which uses a lower cut on the gluon energy $\Delta E$ in the phase-space integration to get finite real 
corrections.\footnote{The implementation of a dedicated dipole subtraction method \`a la Catani-Seymour \cite{Catani-Seymour} is work in 
progress and subject to a later publication.} The divergent part of the phase-space integral can then be performed analytically in the limit of gluon with low energy - the so called soft-gluon approximation. Divergences obtained in the soft-gluon approximation cancel analytically exactly with those coming from the virtual 
corrections. The dependence on this introduced cut should in principal completely vanish. However, in practice the cancellation is limited by the stability of numerical integration of the real corrections. We have verified that in our calculation and {\Fortran}-implementation all cross sections are insensitive to the exact choice of this cut. For a more detailed discussion, we refer the reader to our paper \cite{StopCoannOur1}.
\subsection{Numerical results}
\begin{figure*}[t]
	\includegraphics[scale=0.38]{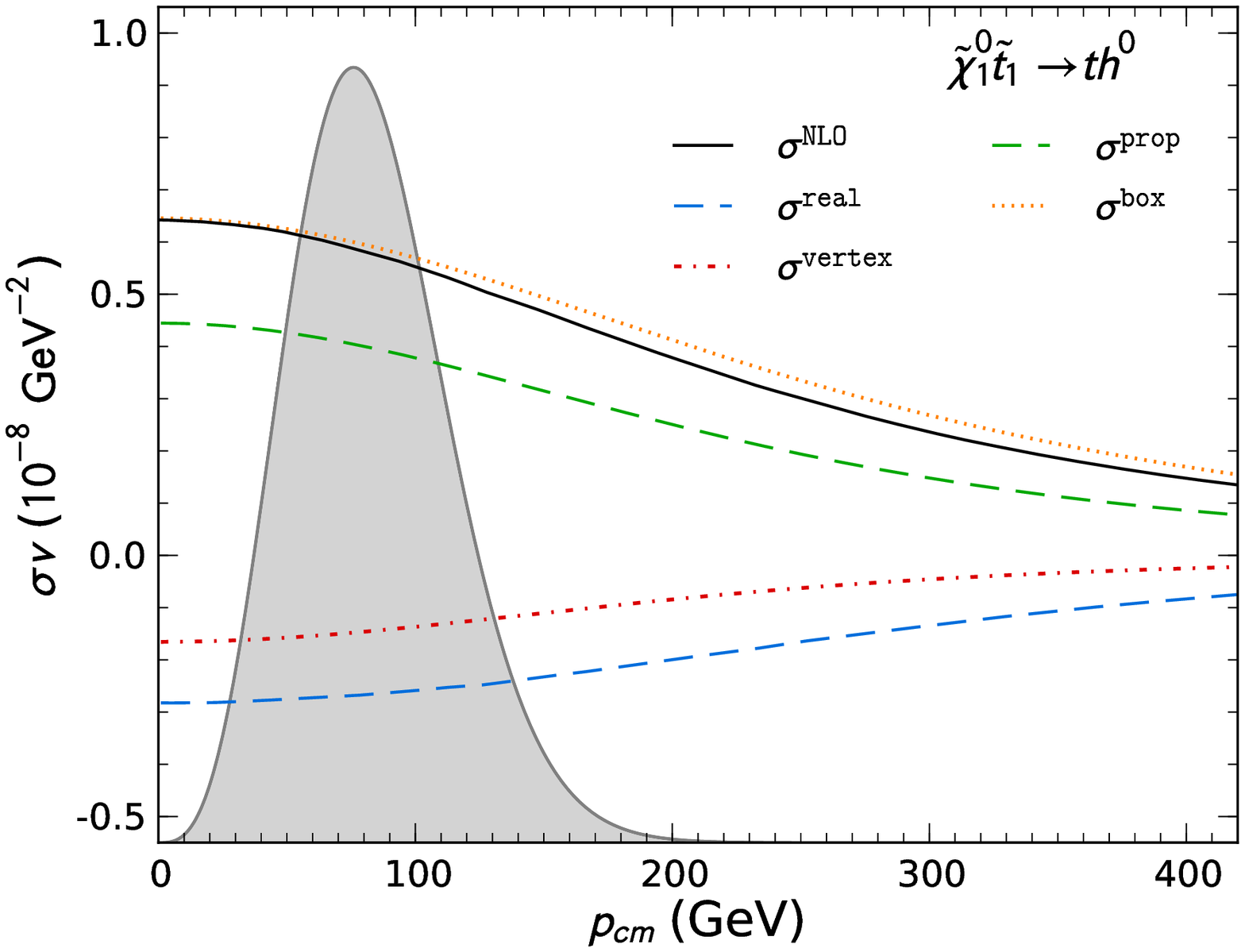}
	\includegraphics[scale=0.38]{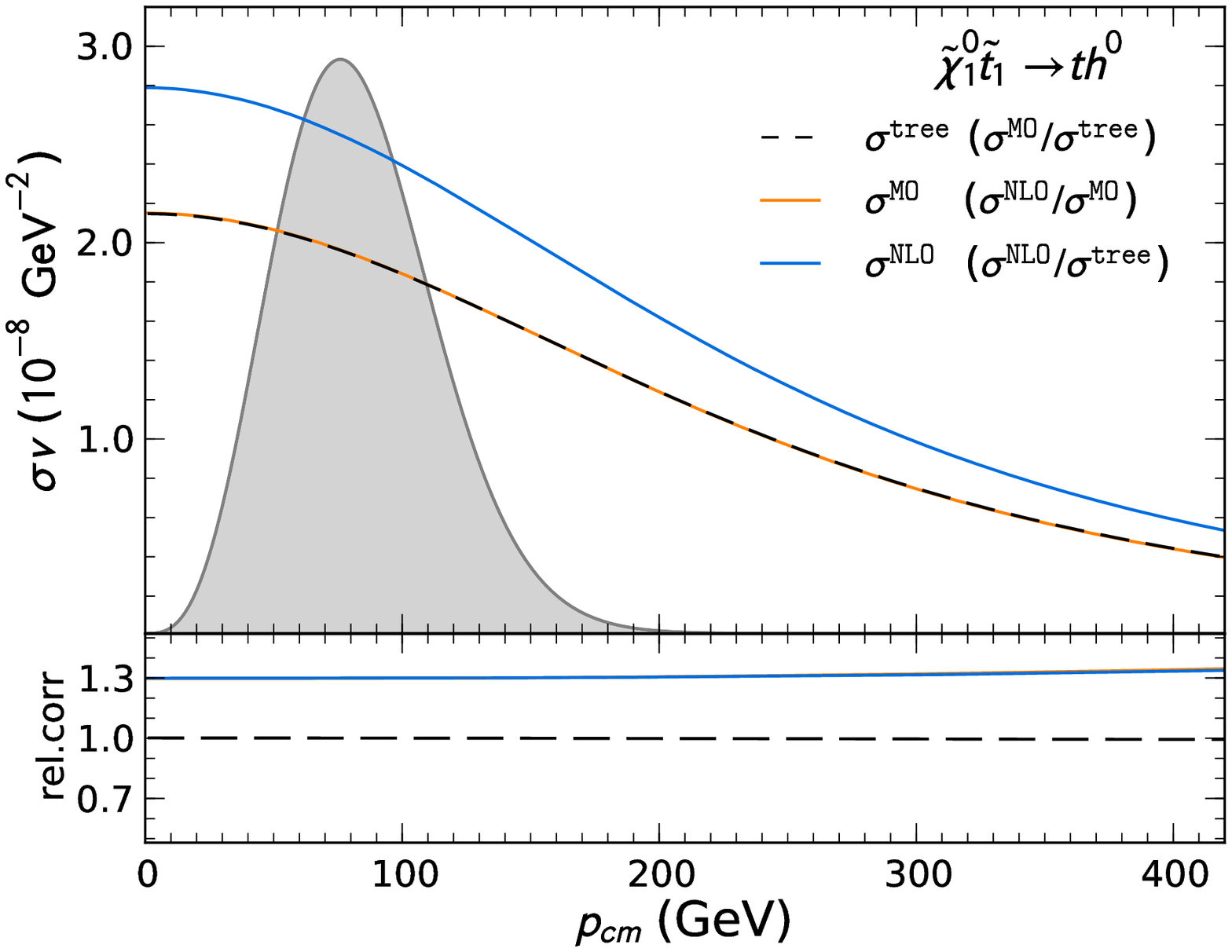}
	\caption{On the left-hand side the contribution of the different corrections to the total next-to-leading order correction for the process $\tilde{\chi}_1^0 \tilde{t}_1 \to t h^0$ is shown. The gray area indicates the thermal distribution (in arbitrary units). On the right-hand side the tree-level (black dashed line), full one-loop (blue solid line) and {\MO} (orange solid line) cross section for the process $\tilde{\chi}_1^0 \tilde{t}_1 \to t h^0$ is depicted. The upper part shows the absolute value of $\sigma v$ together with the thermal distribution (in arbitrary units), whereas the lower part shows the corresponding relative shift (second item in the legend).}
	\label{Fig:NumericalResult}
\end{figure*}

First, we focus on the impact of the calculated next-to-leading order corrections on the neutralino-stop co-annihilation cross section. In Fig.~\ref{Fig:NumericalResult} the next-to-leading correction to the cross section $\sigma v$ (without the tree-level contribution) with its contributions of the different corrections is depicted. Although all renormalized contributions are UV finite, the box, vertex and real part of the correction are IR divergent. 
The IR divergent parts still contain uncancelled poles along with uncancelled logarithms of the large separation scale, which leads to an ambiguity in the exact definition of these contributions. The large remaining logarithms cause the box contribution to be artificially large and leads to negative real corrections at the same time. The latter consists of the soft and the hard part and is thus cut-off independent.\\
In addition one notices that the propagator corrections is significantly enhanced, which holds relatively also for the box contributions. This is due to the fact that this studied process is dominated by the $t$-channel exchange. One of the corrections to the $t$-channel leads to a correction to the stop propagator as well as to a box diagram where a gluon is exchanged between the initial state squark and the final state quark. Thus the enhanced box and propagator corrections entail a large overall next-to-leading order correction in the case of neutralino-stop co-annihilation into a Higgs boson.\\
This is depicted in Fig.~\ref{Fig:NumericalResult}, where a comparison of our tree-level calculation, the effective tree-level calculation of {\MO} and our full one-loop calculation is shown. In the upper panel the corresponding cross section $\sigma v$ is depicted, in the lower part we show the ratio between the different cross sections. As it can be seen, our tree-level calculation is in agreement with the values given by {\MO}. Taking into account our full one-loop calculation the cross section increases by about 30\% in comparision to the tree-level. This is caused by the significant contribution from the box diagrams and propagator corrections as discussed above.
\section{Impact on the relic density}
\begin{figure*}[t]
	\includegraphics[scale=0.39]{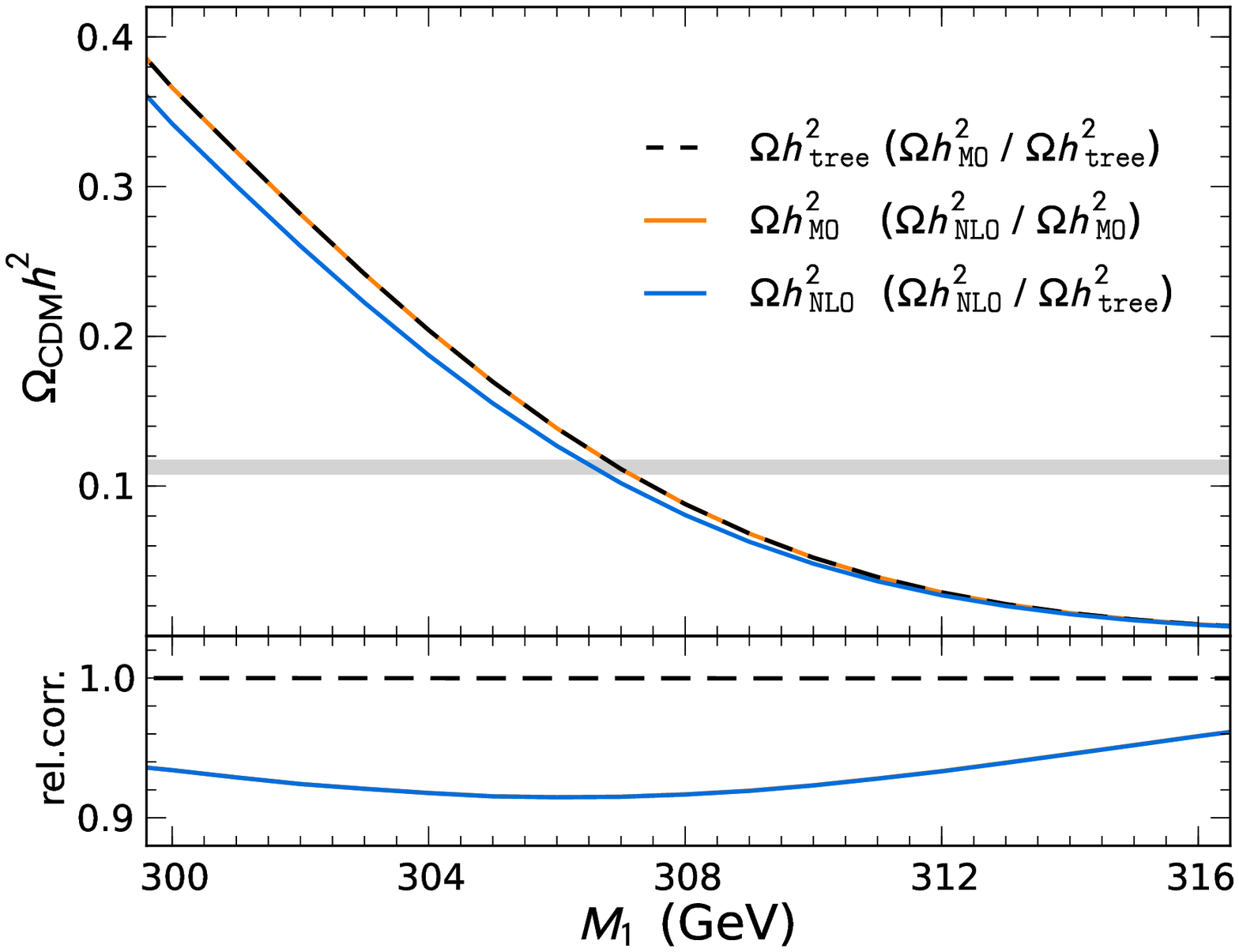}
	\includegraphics[scale=0.39]{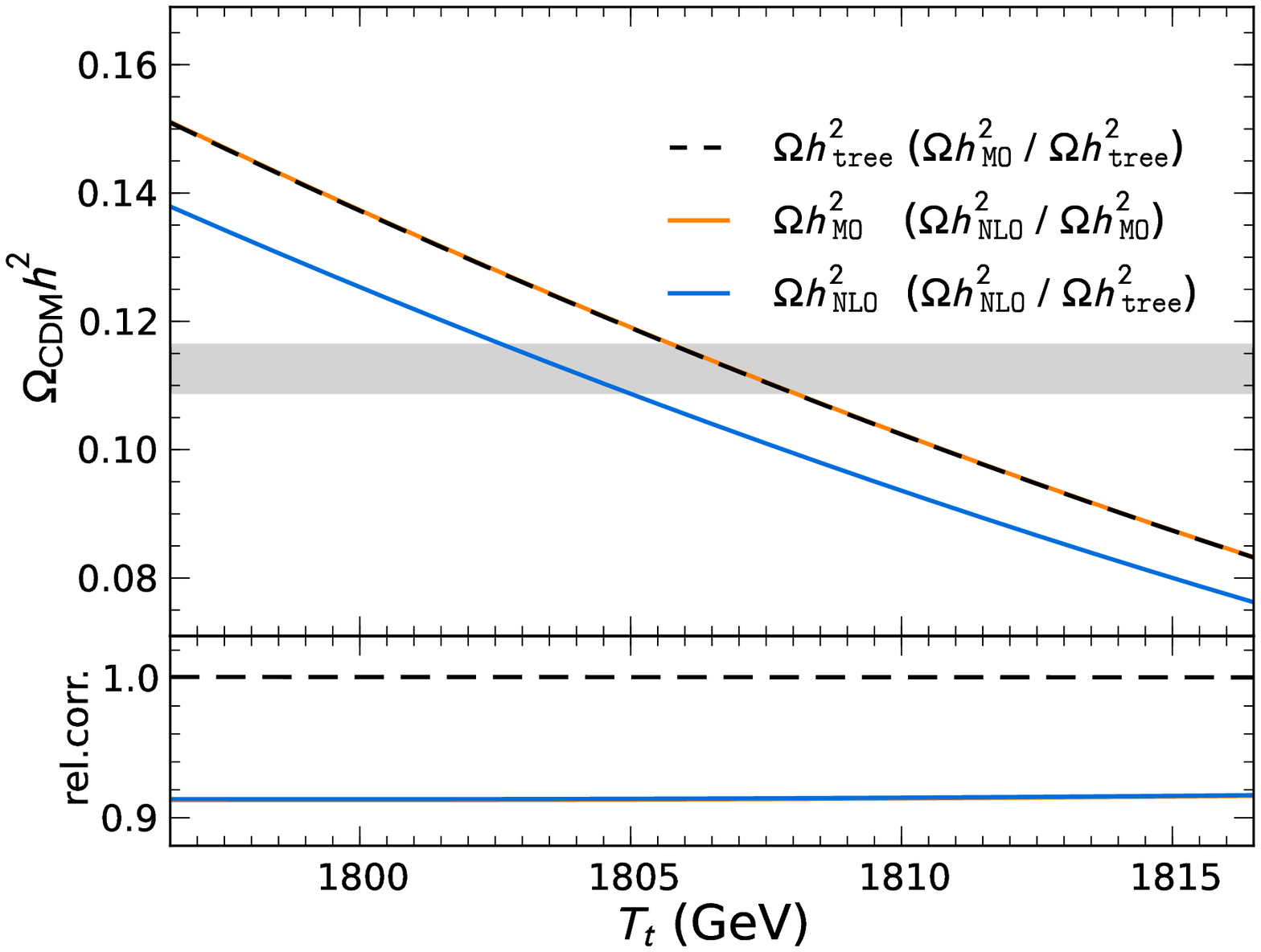}
	\caption{The neutralino relic density $\Omega_{\chi}h^2$ as a function of $M_1$ 
	(left) and $T_t$ (right) calculated using different co-annihilation cross sections: 
	default {\MO} (orange solid line), tree-level (black dashed line), and full one-loop (blue solid line). The gray band indicates the 
	favored range according to Eq.\ (\protect\ref{Eq:WMAP}). The lower part shows the relative impact 
	of the one-loop correction on the relic-density compared to the tree-level calculation.}
	\label{Fig:Relic1DScenario1}
\end{figure*}

Our numerical implementation of the full on-loop calculation, described in the previous section, can be used as an extension to public dark matter tools like {\MO} to evaluate the impact of the one-loop corrections on the neutralino relic density. Even if this study concentrates on the case of co-annihilation of the lightest neutralino with the lightest stop, our chosen implementation is general so that the code can be used for any neutralino-sfermion co-annihilation process.\\
In the following we compare the neutralino relic density obtained by the three in Sec. \ref{Sec:Calculation} discussed calculations: the cross-section calculated by default through {\MO} at tree-level, our tree-level calculation and our full next-to-leading order calculation.\\
First, we study the change of the relic density when varying a single input parameter around the value of the studied scenario of Tab.~\ref{Tab:Scenarios}. In Fig.~\ref{Fig:Relic1DScenario1} the relic density $\Omega_{\chi}h^2$ is shown as a function of the bino mass parameter $M_1$ and the trilinear coupling parameter $T_t$, calculated on the basis of the aforementioned three different calculations.\\
The predicted relic density is very sensitive to variations of the bino mass parameter. It decreases rapidly for higher values of $M_1$ due to a smaller mass splitting between the lightest neutralino and the lightest stop, which enhances the stop-stop annihilation along with the neutralino-stop co-annihilation. In contrast, slightly lower values for the bino mass parameter enlarge the mass difference and suppress the contribution of co-annihilation processes in favor of neutralino-neutralino annihilation. The predicted relic density is then higher due to the absence of co-annihilation.\\
The lower part of Fig.~\ref{Fig:Relic1DScenario1} shows the relative correction, i.e. the ratio of the relic density calculated with our full one-loop co-annihilation cross section to the one included by default in {\MO} and our tree-level, respectively, is shown.
Taking into account the full one-loop calculation, a relative correction to the effective tree-level implemented in {\MO} of about 9\% is observed. This is due to the lightest Higgs final state, which contributes around 38.5\% to the total (co-)annihilation cross section with a corresponding correction of around 30\% (see Fig.~\ref{Fig:NumericalResult}). With the current experimental uncertainty of about 3\% according to Eq.~(\ref{Eq:WMAP}), the impact of the presented corrections is significant and thus important to be taken into account.\\
The relic density is less sensitive to varying the trilinear coupling parameter $T_t$ around the value of the chosen scenario ($T_t=1806.5$ GeV). This is shown on the right-hand side of Fig.~\ref{Fig:Relic1DScenario1}. The difference between the uncorrected and corrected relic density in the cosmologically favored region corresponds to a difference of approximately 3 GeV in $T_t$.\\
\begin{figure*}[t]
	\includegraphics[scale=0.43]{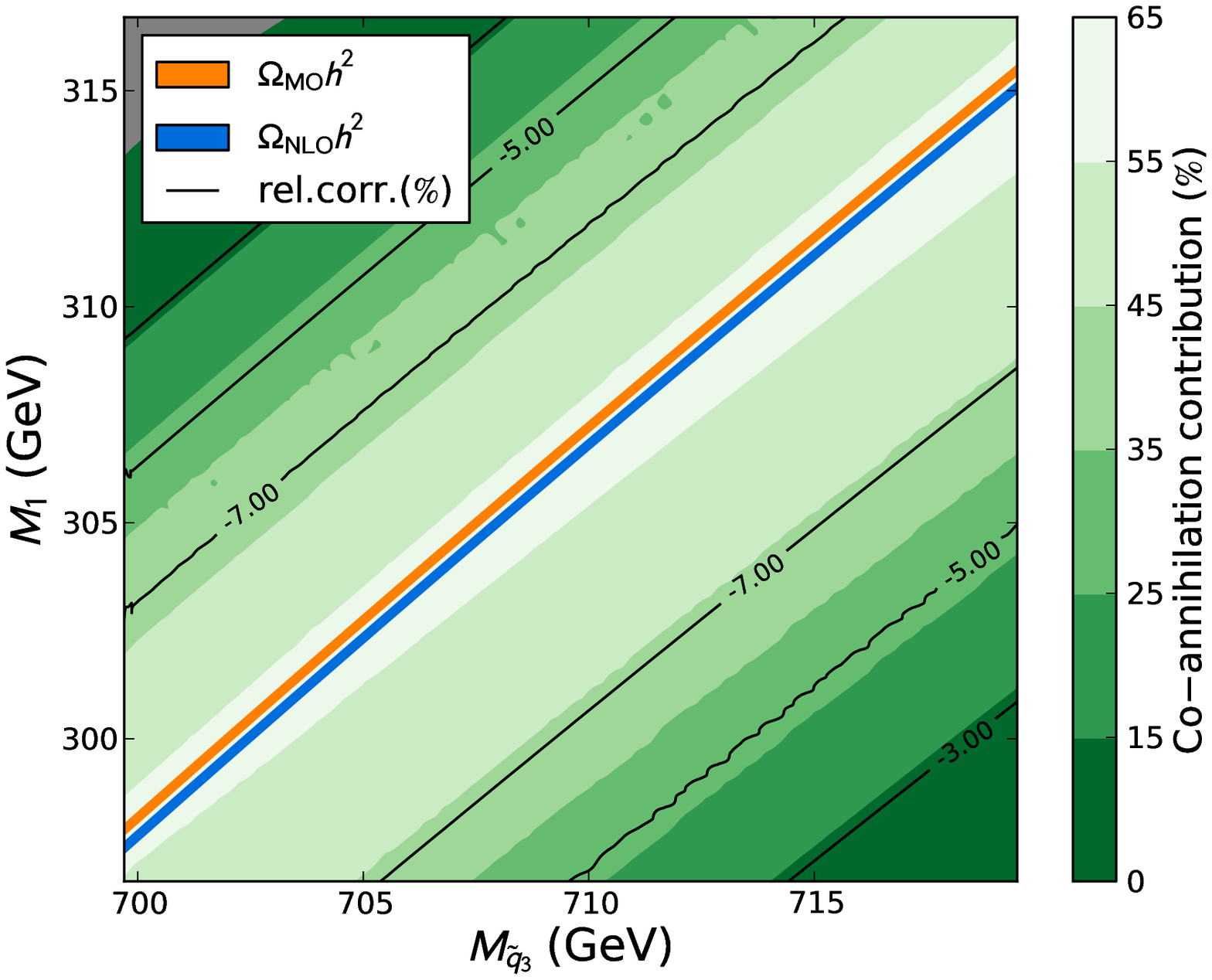}
	\includegraphics[scale=0.43]{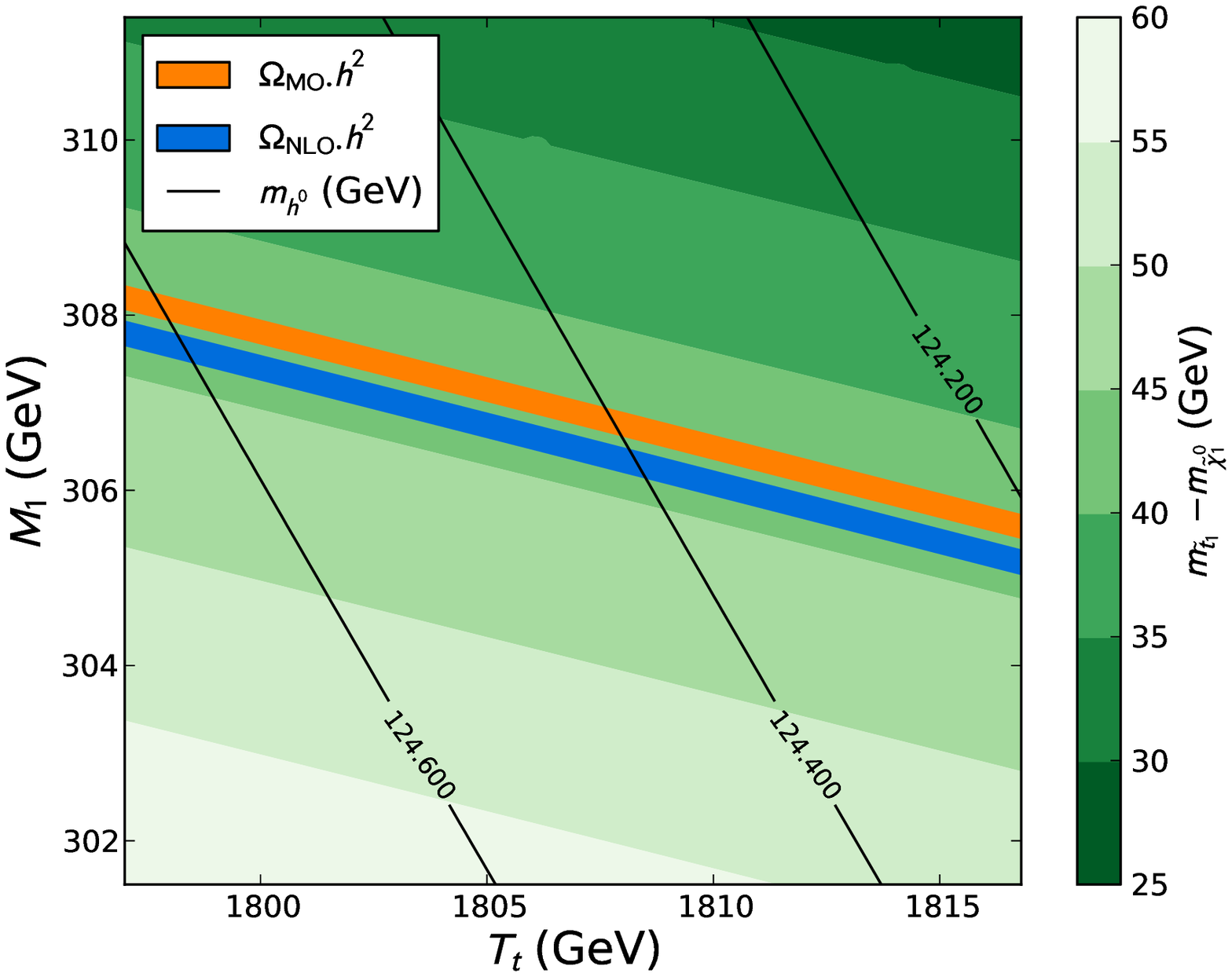}\\
	\caption{WMAP-compatible relic density bands from default {\MO} calculation (orange) and our one-loop calculation for co-annihilation (blue) in the $(M_{\tilde{q}_3},M_1)$ (left) and $(T_t,M_1)$ (right) plane. In the plot on the left hand side the relative contribution of co-annihilation processes is shown in green contour, and the relative impact of the one-loop corrections on the relic density in black lines. The plot on the right hand side shows the LSP-NLSP mass difference in green contour, and the lightest neutral Higgs boson mass in black lines.}
	\label{Fig:Relic2D}
\end{figure*}%
More details on the impact of the full next-to-leading order corrections are shown in Fig.~\ref{Fig:Relic2D}, where the WMAP favored region is depicted as a function of two parameters. On the left hand side we show the dependence on the mass parameter of the third generation of squarks $M_{\tilde{q}_3}$ and the bino mass parameter $M_1$. On the right hand side the dependence of the trilinear coupling parameter $T_t$ and again the bino mass parameter $M_1$ is shown. The WMAP-favored region of parameter space within an $1 \sigma$ interval is marked as a band. In orange we show the cosmologically favored region based on the default {\MO} calculation, in blue based on our full one-loop calculation. A clear separation of the two bands is visible, which indicates that the resulting corrections are larger than the current experimental uncertainties.\\
On the left hand side, the relative impact of the one-loop corrections on the relic density is shown in black lines. As already discussed before, corrections up to 9\% are obtained. Also visible in this plot, the WMAP-favored band follows a straight line in the $M_1$-$M_{\tilde{q}_3}$ plane, which corresponds to a constant mass difference between the lightest neutralino and the lightest stop of about 40 GeV. Above the cosmologically allowed band the neutralino becomes heavier and the mass difference decreases. As a consequence, the stop-stop annihilation becomes dominant. As stop-stop annihilation has typically a significant higher cross section than the co-annihilation processes, this leads to a too small neutralino relic density. For large values of $M_1$ (gray area in the upper left corner) the stop becomes the lightest supersymmetric particle, which is disfavored as dark matter candidate both for its electric and color charge. Below the cosmologially allowed band, the mass difference gets larger and thus the neutralino-stop and stop-stop (co-)annihilation becomes more and more Boltzmann suppressed. In contrast, the neutralino annihilation, which has a lower cross section, becomes now dominant, which results in too large relic density.\\
Focusing on the right hand side of Fig.~\ref{Fig:Relic2D}, we see in different green colors the mass difference between the lightest and next-to-lightest supersymmetric particle depicted. It confirms that the WMAP-favored region follows a contour of a constant mass difference around $40-45~{\rm GeV}$. In this plot the solid black lines show the mass of the lightest Higgs boson. The whole cosmologically favored region lies within the recent Higgs mass limit of $125.2~{\rm GeV} \pm 0.9~{\rm GeV}$ \cite{ATLAS2012update}. At the moment the cosmologically constraints from WMAP are more stringent than the current bounds on a Higgs-like particle.\\
We can conclude that the impact of the studied one-loop corrections on the dark matter relic density is larger than the current experimental uncertainties by WMAP. Therefore it is necessary to take these corrections into account for a solid theoretical prediction of the neutralino relic density.

\section{Conclusions}

A powerful method to constrain the parameter space of theories beyond the Standard Model is to compare the predicted dark matter relic density with data from cosmological precision measurements, in particular from the WMAP satellite. On the particle physics side, one of the main uncertainties on the relic density prediction arises from the calculation of the (co-)annihilation cross sections of the dark matter particle.\\
Therefore we studied the impact of one-loop corrections to neutralino-stop co-annihilation into electroweak vector bosons or Higgs bosons on the neutralino relic density. Especially with the recent developments regarding the discovery of a new boson with a mass of around 126 GeV, which can be interpreted as the lightest Higgs boson within the MSSM, the phenomenology of neutralino-stop co-annihilation is even more motivated. A favored sizeable trilinear coupling parameter $T_t$ increases the relative importance of the neutralino-stop co-annihilation into a top quark and a Higgs boson at the same time. In our presented scenario such a process contributes around 40 \% to the overall (co-)annihilation cross section and receives one-loop corrections of around 30 \%, such that an impact on the neutralino relic density of around 9 \% is visible. As this is larger than the current experimental uncertainty of WMAP, it is necessary to take into account such corrections for the theoretical prediction of the dark matter relic density.\\
For a more detailed discussion of the impact of neutralino-stop co-annihilation in different scenarios we refer to our recent publication \cite{StopCoannOur1}. The neutralino-stop co-annihilation into a gluon will be the subject of a later publication.


\begin{thebibliography}{99}


\bibitem{WMAP7}
  The WMAP collaboration, E.~Komatsu {\it et al.},~\emph{Seven-Year Wilkinson Microwave Anisotropy Probe (WMAP) Observations: Cosmological Interpretation},~\emph{Astrophys.\ J.\ Suppl.\ }~{\bf 192}~(2011)~18~ [{\tt arXiv:1001.4538 [astro-ph.CO]}].


\bibitem{GriestSeckel}
    K.~Griest and D.~Seckel,
    \emph{Three exceptions in the calculation of relic abundances},
    \emph{Phys.\ Rev.\  D} {\bf 43}, (1991) 3191.

\bibitem{EdsjoGondolo}
    J.~Edsjo and P.~Gondolo,
    \emph{Neutralino relic density including coannihilations},
    \emph{Phys.\ Rev.\  D} {\bf 56}, (1997) 1879
    [{\tt arXiv:hep-ph/9704361}].

  
\bibitem{Hamann}
  J.~Hamann, S.~Hannestad, M.~S.~Sloth and Y.~Y.~Y.~Wong,
  \emph{How robust are inflation model and dark matter constraints from cosmological data?},
  \emph{Phys.\ Rev.\ D} {\bf 75} (2007) 023522
  [{\tt arXiv:astro-ph/0611582}].

\bibitem{Arbey}
  A.~Arbey and F.~Mahmoudi,
  \emph{SUSY constraints from relic density: High sensitivity to pre-BBN expansion rate},
  \emph{Phys.\ Lett.\ B} {\bf 669} (2008) 46
  [{\tt arXiv:0803.0741 [hep-ph]}].

\bibitem{Belanger}
  G.~B\'elanger, S.~Kraml and A.~Pukhov,
  \emph{Comparison of SUSY spectrum calculations and impact on the relic density constraints from WMAP},
  \emph{Phys.\ Rev.\ D} {\bf 72} (2005) 015003
  [{\tt arXiv:hep-ph/0502079}].

  


\bibitem{DarkSusy}
  P.~Gondolo, J.~Edsjo, P.~Ullio, L.~Bergstrom, M.~Schelke and E.~A.~Baltz,
  \emph{DarkSUSY: Computing supersymmetric dark matter properties numerically},
  \emph{JCAP} {\bf 0407} (2004) 008
  [{\tt arXiv:astro-ph/0406204}].\\
  P. Gondolo, J. Edsjo, P. Ullio, L. Bergstrom, M. Schelke, E.A. Baltz, T. Bringmann and G. Duda, http://www.darksusy.org

\bibitem{micrOMEGAs2007}
  G.~B\'elanger, F.~Boudjema, A.~Pukhov and A.~Semenov,
  \emph{micrOMEGAs 2.0.7: A program to calculate the relic density of dark matter in a generic model},
  \emph{Comput.\ Phys.\ Commun.}\  {\bf 177} (2007) 894. \\
  G.~B\'elanger, F.~Boudjema, A.~Pukhov and A.~Semenov,
  \emph{MicrOMEGAs: A Program for calculating the relic density in the MSSM},
  \emph{Comput.\ Phys.\ Commun.}\  {\bf 149} (2002) 103
  [{\tt arXiv:hep-ph/0112278}].


\bibitem{DMNLO_AFunnel}
  B.~Herrmann and M.~Klasen,
  \emph{SUSY-QCD Corrections to Dark Matter Annihilation in the Higgs Funnel},
  \emph{Phys.\ Rev.\ D} {\bf 76} (2007) 117704
  [{\tt arXiv:0709.0043 [hep-ph]}].

\bibitem{DMNLO_mSUGRA}
  B.~Herrmann, M.~Klasen and K.~Kovarik,
  \emph{Neutralino Annihilation into Massive Quarks with SUSY-QCD Corrections},
  \emph{Phys.\ Rev.\ D} {\bf 79} (2009) 061701
  [{\tt arXiv:0901.0481 [hep-ph]}].

\bibitem{DMNLO_NUHM}
  B.~Herrmann, M.~Klasen and K.~Kovarik,
  \emph{SUSY-QCD effects on neutralino dark matter annihilation beyond scalar or gaugino mass unification},
  \emph{Phys.\ Rev.\ D} {\bf 80} (2009) 085025
  [{\tt arXiv:0907.0030 [hep-ph]}].
  
  
  

\bibitem{Sloops2007}
  N.~Baro, F.~Boudjema and A.~Semenov,
  \emph{Full one-loop corrections to the relic density in the MSSM: A Few examples},
  \emph{Phys.\ Lett.\ B} {\bf 660} (2008) 550
  [{\tt arXiv:0710.1821 [hep-ph]}].

\bibitem{Sloops2009}
  N.~Baro, G.~Chalons and S.~Hao,
  \emph{Coannihilation with a chargino and gauge boson pair production at one-loop},
  \emph{AIP Conf.\ Proc.\ } {\bf 1200} (2010) 1067
  [{\tt arXiv:0909.3263 [hep-ph]}].

\bibitem{Sloops2010}
  N.~Baro, F.~Boudjema, G.~Chalons and S.~Hao,
  \emph{Relic density at one-loop with gauge boson pair production},
  \emph{Phys.\ Rev.\ D} {\bf 81} (2010) 015005
  [{\tt arXiv:0910.3293 [hep-ph]}].

\bibitem{Sloops2011}
  F.~Boudjema, G.~Drieu La Rochelle and S.~Kulkarni,
  \emph{One-loop corrections, uncertainties and approximations in neutralino annihilations: Examples},
  \emph{Phys.\ Rev.\ D} {\bf 84} (2011) 116001
  [{\tt arXiv:1108.4291 [hep-ph]}].

\bibitem{EffCouplings}
  A.~Chatterjee, M.~Drees and S.~Kulkarni,
  \emph{Radiative Corrections to the Neutralino Dark Matter Relic Density - an Effective Coupling Approach},
  [{\tt arXiv:1209.2328 [hep-ph]}].

\bibitem{Freitas2007}
  A.~Freitas,
  \emph{Radiative corrections to co-annihilation processes},
  \emph{Phys.\ Lett.\ B} {\bf 652} (2007) 280
  [{\tt arXiv:0705.4027 [hep-ph]}].



\bibitem{StopCoann1}
   C.~Boehm, A.~Djouadi and M.~Drees,
   \emph{Light Scalar Top Quarks and Supersymmetric Dark Matter},
   \emph{Phys.\ Rev.\ D} {\bf 62} (2000) 035012
   [{\tt arXiv:hep-ph/9911496}].


\bibitem{StopCoann2}
    J.~Ellis, K.~A.~Olive and Y.~Santoso,
    \emph{Calculations of Neutralino-Stop Coannihilation in the CMSSM},
    \emph{Astropart.\ Phys.}\ {\bf 18} (2003) 395
    [{\tt arXiv:hep-ph/0112113}].

\bibitem{naturalSUSY1}
	  M.~Papucci, J.~T.~Ruderman and A.~Weiler,
  		\emph{Natural SUSY Endures},
  		\emph{JHEP} {\bf 1209} (2012) 035
  		[{\tt arXiv:1110.6926 [hep-ph]}].

\bibitem{naturalSUSY2}
      R.~Auzzi, A.~Giveon, S.~B.~Gudnason and T.~Shacham,
      \emph{A Light Stop with Flavor in Natural SUSY},
      {\tt arXiv:1208.6263 [hep-ph]}.
%
%



\bibitem{ATLAS2012}
    G.~Aad {\it et al.}  [ATLAS Collaboration],
    \emph{Observation of a new particle in the search for the Standard Model Higgs boson with the ATLAS detector at the LHC},
    \emph{Phys.\ Lett.\ B} {\bf 716} (2012) 1
    [{\tt arXiv:1207.7214 [hep-ex]}].

\bibitem{CMS2012}
	  S.~Chatrchyan {\it et al.}  [CMS Collaboration],
	  \emph{Observation of a new boson at a mass of 125 GeV with the CMS experiment at the LHC},
	  \emph{Phys.\ Lett.\ B} {\bf 716} (2012) 30
	  [{\tt arXiv:1207.7235 [hep-ex]}].
	  
\bibitem{ATLAS2012update}
          [ATLAS Collaboration]
          \emph{An update of combined measurements of the new Higgs-like boson with high mass resolution channels},
          ATLAS-CONF-2012-170, Dec, 2012.


\bibitem{Arbey2012}
      A.~Arbey, M.~Battaglia, A.~Djouadi and F.~Mahmoudi,
      \emph{An update on the constraints on the phenomenological MSSM from the new LHC Higgs results},
      {\tt arXiv:1211.4004 [hep-ph]}.

\bibitem{Haber1996}
	H.~E.~Haber, R.~Hempfling and A.~H.~Hoang,
	\emph{Approximating the radiatively corrected Higgs mass in the minimal supersymmetric model},
	\emph{Z.\ Phys.\ C} {\bf 75} (1997) 539
	[{\tt arXiv:hep-ph/9609331}].

\bibitem{Badziak2012}
	M.~Badziak, E.~Dudas, M.~Olechowski and S.~Pokorski,
	\emph{Inverted Sfermion Mass Hierarchy and the Higgs Boson Mass in the MSSM},
	\emph{JHEP} {\bf 1207} (2012) 155
	[{\tt arXiv:1205.1675 [hep-ph]}].
	  

\bibitem{SPA2005}
  J.~A.~Aguilar-Saavedra, A.~Ali, B.~C.~Allanach, R.~L.~Arnowitt, H.~A.~Baer, J.~A.~Bagger, C.~Balazs and V.~D.~Barger {\it et al.},
  \emph{Supersymmetry parameter analysis: SPA convention and project},
  \emph{Eur.\ Phys.\ J.\ C} {\bf 46} (2006) 43
  [{\tt arXiv:hep-ph/0511344}].

  \bibitem{SPheno}
  W.~Porod,
  \emph{SPheno, a program for calculating supersymmetric spectra, SUSY particle decays and SUSY particle production at e+ e- colliders},
  \emph{Comput.\ Phys.\ Commun.}\  {\bf 153} (2003) 275
  [{\tt arXiv:hep-ph/0301101}]. \\
  W.~Porod and F.~Staub,
  \emph{SPheno 3.1: Extensions including flavour, CP-phases and models beyond the MSSM},
  \emph{Comput.\ Phys.\ Commun.}\  {\bf 183} (2012) 2458
  [{\tt arXiv:1104.1573 [hep-ph]}].
	  
\bibitem{PDG2012}
  J.~Beringer {\it et al.}  [Particle Data Group Collaboration],
   \emph{Review of Particle Physics (RPP)},
   \emph{Phys.\ Rev.\ D} {\bf 86} (2012) 010001.


\bibitem{HFAG}
  D.~Asner {\it et al.} [Heavy Flavor Averaging Group Collaboration],
  \emph{Averages of b-hadron, c-hadron, and $\tau$-lepton Properties},
  {\tt arXiv:1010.1589 [hep-ex]}, and online update at {\tt http://www.slac.stanford.edu/xorg/hfag}.



\bibitem{StopCoannOur1}
  J.~Harz, B.~Herrmann, M.~Klasen, K. Kova\v{r}\'{\i}k and Q. Le Boulc'h,
   \emph{Neutralino-stop co-annihilation into electroweak gauge and Higgs bosons at one loop},
   submitted to \emph{Phys.\ Rev.\ D},
   {\tt arXiv:1212.5241 [hep-ph]}.


\bibitem{FeynArts}
      T.~Hahn,
       \emph{Generating Feynman diagrams and amplitudes with FeynArts 3},
      \emph{Comput.\ Phys.\ Commun.}\ {\bf 140} (2001) 418
      [{\tt arXiv:hep-ph/0012260}].  

\bibitem{FeynCalc}
      R.~Mertig, M.~B\"ohm and A.~Denner,
       \emph{Feyn Calc - Computer-algebraic calculation of Feynman amplitudes},
      \emph{Comput.\ Phys.\ Commun.}\ {\bf 64} (1991) 345.

\bibitem{FORM}
      J.~A.~M.~Vermaseren,
       \emph{New features of FORM},
      {\tt arXiv:math-ph/0010025}.

\bibitem{DMNLO}
  {\tt http://dmnlo.hepforge.org}.


\bibitem{GieleGlover}
      W.~T.~Giele and E.~W.~N.~Glover,
      \emph{Higher-order corrections to jet cross sections in e+e- annihilation},
      \emph{Phys.\ Rev.\ D} {\bf 46} (1992) 1980.

\bibitem{HarrisOwens}
  B.~W.~Harris and J.~F.~Owens,
  \emph{The Two cutoff phase space slicing method},
  \emph{Phys.\ Rev.\ D} {\bf 65} (2002) 094032
  [{\tt arXiv:hep-ph/0102128}].

\bibitem{Denner:1991kt}
  A.~Denner,
  \emph{Techniques for calculation of electroweak radiative corrections at the one loop level and results for W physics at LEP-200},
  \emph{Fortsch.\ Phys.}\  {\bf 41} (1993) 307
  [{\tt arXiv:0709.1075 [hep-ph]}].

\bibitem{Catani-Seymour}
      S.~Catani, S.~Dittmaier, M.~H.~Seymour and Z.~Trocsanyi,
      \emph{Nucl.\ Phys.\ B} {\bf 627} (2002) 189
      [{\tt arXiv:hep-ph/0201036}].
	  
\end{thebibliography}
\end{document}